# A Mechanism of Long-Range Order Induced by Random Local Fields: Effective Anisotropy Created by Defects

*A.A. Berzin*[1], *A.I. Morosov*[2*], *and A.S. Sigov*[1]

[1] Moscow Technological University (MIREA), 78 Vernadskiy Ave., 119454 Moscow, Russian Federation
[2] Moscow Institute of Physics and Technology (State University), 9 Institutskiy per., 141700 Dolgoprudny, Moscow Region, Russian Federation

## Abstract

We propose a microscopic mechanism of the long-range order in two-dimensional space induced by random local fields of crystal defects. The anisotropic distribution of defect-induced random local field directions in the *n*-dimensional space of vector order parameter with the *O(n)* symmetry is shown to give rise to the defect-induced effective anisotropy in the system. The expression for the effective anisotropy constant is found. A weak anisotropy of the "easy axis" type transforms the *X-Y* model and the Heisenberg model to the class of Ising models, and brings into existence the long-range order in the system.

[*] E-mail: mor-alexandr@yandex.ru

## 1. Introduction

Minchau and Pelcovits [1] discovered the phenomenon of initiation of the long range order in the two-dimensional *X-Y* model at finite temperature resulting from the action of collinear to each other local fields of the “random local field” type defects. The direction of the magnetization arising was perpendicular to the direction of the local fields. Since a long-range order is absent in a pure system at finite temperature, and the Berezinskii-Kosterlitz-Thouless phase takes place [2, 3], this phenomenon has been further named the "random fields induced order" (RFIO) [4]. In Ref. [4] this phenomenon was generalized to the Heisenberg model. As the reason for RFIO occurrence, the violation of the continuous symmetry of the system was indicated, but the microscopic mechanism of RFIO has not been found yet.

It was shown in our preceding paper [5] that for space dimensionality $2<d<4$, the anisotropic distribution of random local fields in the $n$-dimensional space of the order parameter gave rise to the effective anisotropy term (in the second order perturbation theory in random field). Such anisotropy tended to align the order parameter perpendicularly to the preferential direction of random local fields.

In the present paper we demonstrate that a similar effect takes place in the two-dimensional system as well. The specific feature of two-dimensional models, as distinct from those considered in paper [5], is the lack of the long-range order in a perfect system at finite temperature. Hence we must suppose the existence of the long-range order initiated by the random fields and solve the self-consistent problem.

## 2. Energy of the system of classical spins

The exchange interaction energy of $n$-component localized spins $\boldsymbol{S}_i$ comprising the two-dimensional lattice has the form

$$W_{ex} = -\sum_{i,j>i} J_{ij} \boldsymbol{S}_i \boldsymbol{S}_j, \qquad (1)$$

where $J_{ij}$ is the exchange integral for $i$-th и $j$-th spins, and the summation is performed over the whole spin lattice.

The energy of interaction between the spins and defect-induced random local fields is

$$W_{def} = -\sum_l \boldsymbol{S}_l \boldsymbol{h}_l, \tag{2}$$

where the summation is performed over defects randomly located in the lattice sites, and the density of random local fields $\boldsymbol{h}$ distribution in the spin space (order parameter space) possesses the property $\rho(\boldsymbol{h}) = \rho(-\boldsymbol{h})$, which provides the lack of mean field in an infinite system.

Assuming the presence of the uniaxial anisotropy in the system, we represent its energy in the form

$$W_{an} = \frac{1}{2} K_{eff} \sum_l (S_l^z)^2, \tag{3}$$

where $K_{eff}$ is the effective anisotropy constant and $S_i^z$ is the $i$-th spin projection onto the given $z$-axis.

Switching to the continuous distribution of the order parameter $\boldsymbol{\eta}$, we introduce the inhomogeneous exchange energy in the form [6]

$$\widetilde{W}_{ex} = -\frac{1}{2} \int d^2\, \boldsymbol{r} D \frac{\partial \eta^\perp}{\partial x_i} \frac{\partial \eta^\perp}{\partial x_i}, \qquad (4),$$

where $D \sim Jb^4$, $b$ is the interatomic distance, $J$ is the exchange integral reflecting the interaction of the nearest neighbors, and $\boldsymbol{\eta}^\perp(\boldsymbol{r})$ is the order parameter $\boldsymbol{\eta} \sim \boldsymbol{S}_l b^{-2}$ component caused by the random local field and orthogonal to the direction of the average order parameter $\boldsymbol{\eta}_0$.

The energy of the random field $\boldsymbol{h}(\boldsymbol{r})$ interaction with the order parameter $\boldsymbol{\eta}(\boldsymbol{r})$ is

$$W_{def} = -\int d^2\boldsymbol{r}\, \boldsymbol{h}(\boldsymbol{r}) \boldsymbol{\eta}(\boldsymbol{r}), \tag{5}$$

where

$$\boldsymbol{h}(\boldsymbol{r}) = b^2 \sum_l \boldsymbol{h}_l\, \delta(\boldsymbol{r} - \boldsymbol{r}_l). \tag{6}$$

With account of the notations introduced above, the anisotropy energy (3) takes the form

$$W_{an} = \frac{1}{2} K_{eff} b^2 \int d^2\boldsymbol{r}\, (\eta^z(\boldsymbol{r}))^2. \tag{7}$$

## 3. Effective anisotropy

For a qualitative explanation of the cause for the effective anisotropy let us consider the influence of defect-induced local field $\boldsymbol{h}_l$ upon uniform distribution of the order parameter. For simplicity, we neglect the longitudinal susceptibility of the system at low temperatures, much smaller than the temperature of magnetic ordering.

The component of the random field $\boldsymbol{h}_l^{\perp}$ perpendicular to the $\boldsymbol{\eta}_0$ direction leads to a local deviation of the order parameter and to the appearance of the orthogonal component $\boldsymbol{\eta}^{\perp}(\boldsymbol{r})$. The result is a negative additive to the energy of the ground state proportional to $(\boldsymbol{h}_l^{\perp})^2$. It is maximum in modulus when the $\boldsymbol{\eta}_0$ direction is perpendicular to the defect-induced local field.

Let us find the expression for the anisotropy energy in the case of arbitrary distribution of the directions of the defect-induced random fields. We represent the order parameter in the linear in $\boldsymbol{h}$ approximation in the form

$$\boldsymbol{\eta}(\boldsymbol{r}) = \boldsymbol{\eta}_0 + \boldsymbol{\eta}^{\perp}(\boldsymbol{r}), \tag{8}$$

where $|\boldsymbol{\eta}_0| \gg |\boldsymbol{\eta}^{\perp}(\boldsymbol{r})|$. The term proportional to $\boldsymbol{\eta}_0$ in the expression for $W_{def}$ vanishes because of the function $\rho(\boldsymbol{h})$ parity .

The quantity $\boldsymbol{h}^{\perp}(\boldsymbol{r})$ may be represented by a sum

$$\boldsymbol{h}^{\perp}(\boldsymbol{r}) = b^2 \sum_l [\boldsymbol{h}_l - \boldsymbol{n}(\boldsymbol{n}\boldsymbol{h}_l)]\, \delta(\boldsymbol{r} - \boldsymbol{r}_l), \tag{9}$$

where $\boldsymbol{n} = \boldsymbol{\eta}_0/|\boldsymbol{\eta}_0|$.

The Fourier component of the function $\boldsymbol{\eta}^{\perp}(\boldsymbol{k})$ is related to the Fourier component of the random field $\boldsymbol{h}^{\perp}(\boldsymbol{k})$:

$$\boldsymbol{\eta}^{\perp}(\boldsymbol{k}) = \chi^{\perp}(\boldsymbol{k}) \boldsymbol{h}^{\perp}(\boldsymbol{k}), \tag{10}$$

where

$$\chi^{\perp}(\boldsymbol{k}) = \left(Dk^2 + \left|K_{eff}\right|b^2\right)^{-1}. \tag{11}$$

In the Heisenberg model, the "easy plane" type anisotropy is induced by local fields perpendicular to this plane, therefore we need to use the susceptibility component corresponding to the direction perpendicular to both magnetization vector and the easy plane.

The quantity $\boldsymbol{h}^{\perp}(\boldsymbol{k})$ is given by the expression

$$\boldsymbol{h}^{\perp}(\boldsymbol{k}) = \frac{1}{V}\int d^2\,\boldsymbol{r}\boldsymbol{h}^{\perp}(\boldsymbol{r})\exp(-i\boldsymbol{k}\boldsymbol{r}) = \frac{1}{N}\sum_l [\boldsymbol{h}_l - \boldsymbol{n}(\boldsymbol{n}\boldsymbol{h}_l)]\exp(-i\boldsymbol{k}\boldsymbol{r}_l), \tag{12}$$

where $N$ is the number of elementary cells. Then

$$\boldsymbol{\eta}^{\perp}(\boldsymbol{r}) = \frac{1}{N}\sum_{\boldsymbol{k}}\chi^{\perp}(\boldsymbol{k})\sum_l [\boldsymbol{h}_l - \boldsymbol{n}(\boldsymbol{n}\boldsymbol{h}_l)]\exp[i\boldsymbol{k}(\boldsymbol{r} - \boldsymbol{r}_l)], \tag{13}$$

and the energy $W_{def}$ (5) takes the form

$$W_{def} = -\frac{1}{N^2}\int d^2\,\boldsymbol{r}\sum_{\boldsymbol{k},\boldsymbol{k}'}\chi^{\perp}(\boldsymbol{k})\sum_{l,m}[\boldsymbol{h}_l - \boldsymbol{n}(\boldsymbol{n}\boldsymbol{h}_l)]\,[\boldsymbol{h}_m - \boldsymbol{n}(\boldsymbol{n}\boldsymbol{h}_m)]\times$$

$$\times\exp[i\boldsymbol{k}(\boldsymbol{r} - \boldsymbol{r}_l) + i\boldsymbol{k}'(\boldsymbol{r} - \boldsymbol{r}_{\boldsymbol{m}})]. \tag{14}$$

The integration produces the factor $V\delta_{\boldsymbol{k},-\boldsymbol{k}'}$. Ultimately we have

$$W_{def} = -\frac{V}{N^2}\sum_{\boldsymbol{k}}\chi^{\perp}(\boldsymbol{k})\sum_{l,m}[\boldsymbol{h}_l - \boldsymbol{n}(\boldsymbol{n}\boldsymbol{h}_l)]\,[\boldsymbol{h}_m - \boldsymbol{n}(\boldsymbol{n}\boldsymbol{h}_m)]\times$$

$$\times\exp[i\boldsymbol{k}(\boldsymbol{r}_{\boldsymbol{m}} - \boldsymbol{r}_l)]. \tag{15}$$

Due to random distribution of defects in the coordinate space, the contribution different from zero results from the summands with $l = m$. Therefore Eq. (15) yields

$$W_{def} = -\frac{V}{N^2}\sum_{\boldsymbol{k}}\chi^{\perp}(\boldsymbol{k})\sum_l [\boldsymbol{h}_l - \boldsymbol{n}(\boldsymbol{n}\boldsymbol{h}_l)]^2 =$$

$$= -xb^2\sum_{\boldsymbol{k}}\chi^{\perp}(\boldsymbol{k})\,\langle[\boldsymbol{h}_l - \boldsymbol{n}(\boldsymbol{n}\boldsymbol{h}_l)]^2\rangle. \tag{16}$$

Here $x$ is the dimensionless concentration of defects (the number of defects per a unit cell), and the brackets $\langle\rangle$ denote averaging over all defect local fields. Going from summation over $\boldsymbol{k}$ to integration over the Brillouin zone and introducing the notation

$$\tilde{\chi}^{\perp} = b^2 \int \frac{d^2\boldsymbol{k}}{(2\pi)^2} \chi^{\perp}(\boldsymbol{k}) = \frac{1}{4\pi b^2 J} \ln \frac{4\pi J}{K_{eff}}, \tag{17}$$

we get the volume density of the energy of interaction between the order parameter and defect-induced random local fields

$$w_{def} = -x\tilde{\chi}^{\perp}\left[\langle \boldsymbol{h}_l{}^2 \rangle - \langle (\boldsymbol{n}\boldsymbol{h}_l)^2 \rangle\right]. \tag{18}$$

One can readily see that in the case of anisotropic distribution of random field directions, the second summand in square brackets in the right-hand side of Eq. (18) describes anisotropy in the order parameter space.

In particular, for the collinear orientation of random fields, the volume density of anisotropy energy takes the form

$$w_{an} = x\tilde{\chi}^{\perp}\langle \boldsymbol{h}_l{}^2 \rangle \cos^2\varphi \equiv \frac{1}{2} K_{eff} S^2 b^{-2} \cos^2\varphi, \tag{19}$$

where $\varphi$ is the angle between the order parameter vector and the axis of "hard magnetization" (defect-induced random local fields are collinear to this axis), and $S$ is the spin vector magnitude.

The self-consistency equation for the effective anisotropy constant can be obtained from Eq. (19):

$$K_{eff} = \frac{x\langle \boldsymbol{h}_l{}^2 \rangle}{2\pi J S^2} \ln \frac{4\pi J}{K_{eff}}. \tag{20}$$

Within the logarithmic approximation, Eq. (20) gives

$$K_{eff} = \frac{x\langle \boldsymbol{h}_l{}^2 \rangle}{2\pi J S^2} \ln \frac{8\pi^2 J^2 S^2}{x\langle \boldsymbol{h}_l{}^2 \rangle}. \tag{21}$$

In the case of coplanar and isotropic in the selected plane distribution of random fields in the Heisenberg model the anisotropy energy volume density is

$$w_{an} = -\frac{1}{2} x\tilde{\chi}^{\perp}\langle \boldsymbol{h}_l{}^2 \rangle \cos^2\varphi \equiv \frac{1}{2} K_{eff} S^2 b^{-2} \cos^2\varphi, \tag{22}$$

$\varphi$ being the angle between the order parameter vector and normal to the plane containing random field vectors. In this instance the effective anisotropy constant is negative and the self-consistency equation looks like

$$|K_{eff}| = \frac{x\langle \boldsymbol{h}_l{}^2 \rangle}{4\pi J S^2} \ln \frac{4\pi J}{|K_{eff}|}, \tag{23}$$

whereas the logarithmic approximation produces

$$|K_{eff}| = \frac{x\langle \boldsymbol{h}_l{}^2 \rangle}{4\pi J S^2} \ln \frac{16\pi^2 J^2 S^2}{x\langle \boldsymbol{h}_l{}^2 \rangle}. \tag{24}$$

For small concentration of defects, the effective anisotropy values deduced above exceed the critical value [7]

$$K_{cr} \sim \frac{x\langle \boldsymbol{h}_l{}^2 \rangle}{J S^2}. \tag{25}$$

By this is meant that the Imry-Ma inhomogeneous state [8] cannot arise. Indeed, to follow the fluctuations of the random field, the order parameter has to deviate from the most favorable (from the point of view of the anisotropy energy) direction. This leads to an increase in the anisotropy energy. When such a growth is not compensated by the gain in energy due to the order parameter alignment with the fluctuations of the random field, the Imry-Ma inhomogeneous state becomes energetically unfavorable.

In the general case of anisotropic distribution of random field directions, one should describe the given anisotropy through the difference $\Delta$ between the maximum and minimum values of the expression $\langle (\boldsymbol{n}, \boldsymbol{h}_l)^2 \rangle$ as a function of the vector $\boldsymbol{n}$ direction. The resulting expressions differ from Eqs. (20) and (21) by substitution of $\Delta$ for the quantity $\langle \boldsymbol{h}_l{}^2 \rangle$.

## 4. Conclusions

It follows from Eq. (19) that for the *X-Y* model the collinear orientation of random local fields induces the “easy axis” type anisotropy, the easy axis being perpendicular to the random field direction. A weak anisotropy of the "easy axis" type transforms the *X*-*Y* model to the class of Ising models [9], that explains the appearance of the long-range order at finite temperature [1]. For the Heisenberg model such an orientation of random local fields induces the “easy plane” type anisotropy. As the result, the Heisenberg model is transformed to the class of *X*-*Y* models, and thus the Berezinskii-Kosterlitz-Thouless transition occurs in the system [9].

In the case of coplanar distribution of random local field directions in the order parameter space in the Heisenberg model, there arises an easy axis perpendicular to the indicated plane. So the Heisenberg model should be transformed to the class of Ising models and the long-range ordering is easily understood [4].

Generally it is possible to state with assurance that behavior of the system with the anisotropy induced by the “random local field” type defects is equivalent to behavior of the perfect system with the same weak anisotropy if the anisotropy constant magnitude exceeds its critical value (25). In the opposite case, the inhomogeneous Imry-Ma state arises in the system.